\documentclass[a4paper,11pt]{article}
\usepackage{jinstpub} 
\usepackage{lineno}
\usepackage{subfigure}
\usepackage{siunitx}
\usepackage{float}
\usepackage{soul}
\usepackage{enumitem}
\sethlcolor{yellow}

\title{\boldmath Simulation Study on the Discrimination of $0\nu\beta\beta$ Events from Single-Electron Events Using Orthogonal-Strip HPGe Detectors}

\author{Qiuli Zhang\textsuperscript{a}, Wenhan Dai\textsuperscript{a}, Peng Zhang\textsuperscript{a}, Mingxin Yang\textsuperscript{a}, Yang Tian\textsuperscript{a}, Zhi Zeng\textsuperscript{a*},Yulan Li\textsuperscript{a}, Ming Zeng\textsuperscript{a}, Hao Ma\textsuperscript{a}, Jianping Cheng\textsuperscript{b}}

\affiliation{\textsuperscript{a}Key Laboratory of Particle and Radiation Imaging (Ministry of Education) and Department of Engineering Physics, Tsinghua University, Beijing 100084, China}
\affiliation{\textsuperscript{b}School of Physics and Astronomy, Beijing Normal University, Beijing 100084, China}

\emailAdd{zengzhi@tsinghua.edu.cn}

\abstract{Neutrinoless double beta decay ($0\nu\beta\beta$) offers a sensitive probe of neutrino mass and its Majorana nature. Orthogonal-strip high-purity germanium (HPGe) detectors with high spatial resolution provide a promising approach for distinguishing $0\nu\beta\beta$ events from single-electron backgrounds. In this work, a simulation framework was developed to evaluate the discrimination performance of these detectors. The framework combined Geant4 simulations with a hybrid numerical-analytical approach to model charge cloud dynamics. A dual-branch convolutional neural network (CNN) was implemented to extract topological features for event classification. The impact of detector geometry on discrimination performance was quantitatively assessed. For a fixed crystal thickness of 15 mm, the background rejection efficiency decreased from 79.5\% to 59.0\% as the strip pitch increased from 0.1 mm to 0.5 mm. For a strip pitch of 0.25 mm, a crystal thickness of 20 mm was found to be optimal, balancing full-energy peak (FEP) efficiency with discrimination capability. These results demonstrate that orthogonal-strip HPGe detectors can  effectively suppress single-electron backgrounds, and provide quantitative guidance for detector design in $^{76}$Ge $0\nu\beta\beta$ decay searches.}

\keywords{HPGe detectors, Neutrinoless double beta decay, Background suppression, Monte Carlo simulation, Charge cloud dynamics}

\begin{document}
\maketitle

\section{Introduction}
\label{sec1}
Neutrinoless double beta decay ($0\nu\beta\beta$) is a critical probe for determining the absolute scale of neutrino mass and confirming the Majorana nature of neutrinos~\cite{0vbb_review}. $^{76}$Ge is a compelling candidate isotope, utilized by experiments such as GERDA, MAJORANA, and CDEX~\cite{gerda_0vbb_2020,majorana_0vbb_2023,cdex_0vbb_Zhang_2024}, owing to the mature technology of high-purity germanium (HPGe) detectors, which offer excellent energy resolution and low intrinsic backgrounds. However, the rarity of $0\nu\beta\beta$ events necessitates ultra-low background conditions. A significant challenge in background suppression arises from single-electron events~\cite{single_electron_czt_zeng,single_electron_czt_bloxham,single_electron_tpc}, which originate from internal cosmogenic radioisotope beta decays or external gamma-ray Compton scattering. These events deposit energy around the $^{76}$Ge $Q$-value (2039~keV), and their track lengths are comparable to those of the $0\nu\beta\beta$ signal.

In conventional single-electrode HPGe detectors, background suppression typically relies on pulse shape analysis (PSA), such as the A/E method~\cite{psa_gerda,psa_majorana}, which effectively distinguishes single-site events (SSEs) from multi-site events (MSEs). However, both $0\nu\beta\beta$ and single-electron events are typically classified as SSEs due to the lack of spatial resolution. Theoretically, a $0\nu\beta\beta$ event emits two electrons, producing a track topology with two Bragg peaks at the endpoints, while a single-electron event produces a single track with only one Bragg peak. Exploiting these topological differences for background rejection requires detectors with high spatial resolution, which is not available in single-electrode HPGe detectors.

Orthogonal-strip HPGe detectors~\cite{amman_germanium_2000,strip_250um_2005,mircostrip_2007}, featuring perpendicular strip electrodes on opposite surfaces, have demonstrated good energy and spatial resolution in fields such as gamma-ray astronomy and medical imaging~\cite{cosi_2020,pet_strip_hpge}. However, their application in $0\nu\beta\beta$ searches remains unexplored. The precise two-dimensional position information provided by fine-pitched strip electrodes offers significant potential for distinguishing $0\nu\beta\beta$ events from single-electron events based on their track topologies. 

The discrimination performance of orthogonal-strip HPGe detectors, however, is fundamentally limited by charge cloud broadening. As carriers drift toward the electrodes, the initial charge cloud expands due to thermal diffusion and Coulomb repulsion, potentially blurring the topological features of the original tracks. Two primary approaches exist for modeling charge cloud dynamics: analytical calculations~\cite{GATTI-cloud,BOGGS-cloud} and full numerical simulations. Analytical models are computationally fast but fail to account for Coulomb repulsion between distinct ionization clusters along the electron track, which becomes significant for MeV-scale events. Full numerical simulations using tools such as SolidStateDetectors.jl~\cite{ssd_simulation_2021} can incorporate drift, diffusion, and Coulomb repulsion, but are computationally prohibitive for MeV-scale $0\nu\beta\beta$ events.

In this work, we developed a hybrid numerical-analytical approach to model charge cloud dynamics and investigated the topological features of $0\nu\beta\beta$ and single-electron events in orthogonal-strip HPGe detectors. Based on these topological differences, a dual-branch convolutional neural network (CNN)~\cite{CNN_2025} was employed to discriminate between these events. Furthermore, we quantitatively assessed the impact of detector geometry on discrimination performance. The remainder of this paper is organized as follows: Section 2 describes the simulation framework, Section 3 outlines the event discrimination methodology, Section 4 presents the results and discusses the impact of detector geometry, and Section 5 draws final conclusions.

\section{Monte Carlo Simulation Framework}
\label{sec2}
\subsection{Basic Simulation Setup}
\label{sec2.1}
\begin{figure}[htbp]
    \centering
    \includegraphics[width=0.6\textwidth]{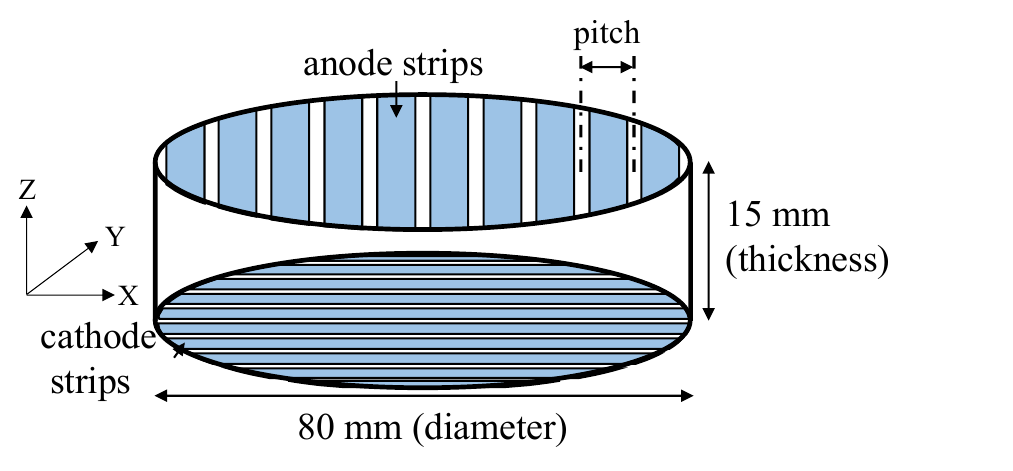}
    \caption{Schematic diagram of the orthogonal-strip HPGe detector. The crystal features an 80-mm diameter and a 15-mm thickness, with orthogonal strip electrodes fabricated on the top and bottom surfaces. The number of strips and their pitch are illustrative and not to scale.}
    \label{fig:detector_geometry}
\end{figure}

The orthogonal-strip HPGe detector investigated in this study consists of a cylindrical crystal with a diameter of 80~mm and a thickness of 15~mm, as illustrated in Fig.~\ref{fig:detector_geometry}. The electrodes on the top and bottom surfaces are segmented into orthogonal strips with a 0.25-mm pitch, providing position sensitivity in the $X$ and $Y$ directions, respectively. In the simulation, each strip was assumed to collect charge over the pitch width.

Two key parameters of the detector geometry were investigated to evaluate their influence on discrimination performance: crystal thickness and strip pitch. The crystal thickness was varied from 5~mm to 40~mm in 5~mm steps. The lower bound was chosen to ensure adequate full-energy-peak (FEP) efficiency, while the upper bound was set by the requirement for full depletion. For each thickness, the bias voltage was adjusted to maintain a constant average electric field of approximately 800~V/cm. To distinguish the sub-millimeter topological differences between $0\nu\beta\beta$ and single-electron events, strip pitches of 0.1~mm, 0.25~mm, and 0.5~mm were investigated. The 0.25-mm pitch represents the high-resolution capabilities demonstrated in the literature~\cite{strip_250um_2005}, while the 0.1-mm and 0.5-mm~\cite{berkeley_ieee_2025} pitches represent an idealized fine-segmentation scenario and a more practical fabrication option, respectively.

\begin{figure}[htbp]
    \centering
    \includegraphics[width=0.75\textwidth]{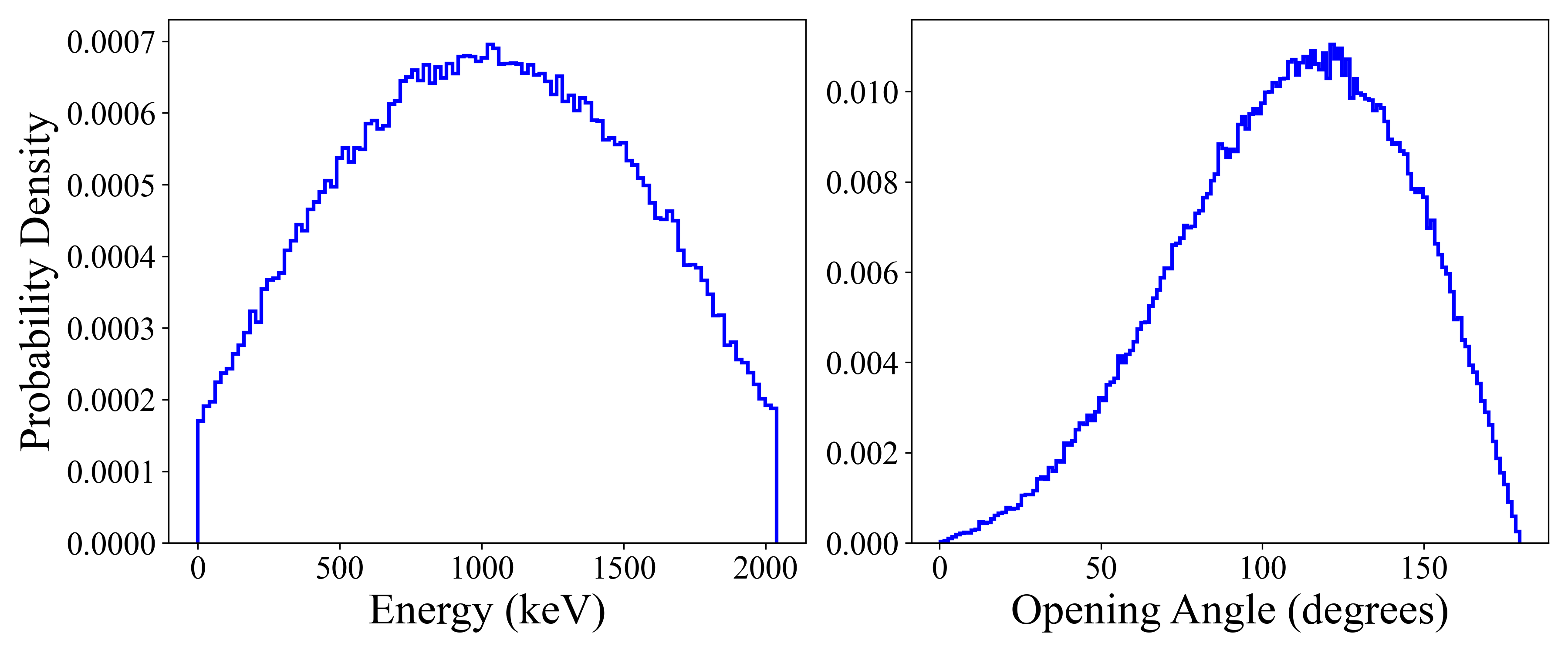}
    \caption{Simulated energy spectrum of individual electrons (left) and the angular distribution of the two electrons (right) for $0\nu\beta\beta$ events generated by BxDecay0.}
    \label{fig:kinematics}
\end{figure}

Signal events ($0\nu\beta\beta$ decay of $^{76}$Ge) were generated using the BxDecay0 package~\cite{bxdecay0}. As illustrated in Fig.~\ref{fig:kinematics}, the simulated energy spectrum of the individual electrons follows a bell-shaped profile peaking at approximately half of 2039~keV. The angular distribution of the two electrons, characterized by their opening angle $\theta$, peaks at $\theta = 120^\circ$~\cite{tables_betadecay,single_electron_czt_bloxham}, indicating a preference for back-to-back emission. For background simulations, single-electron events were generated with an initial kinetic energy of 2039~keV. Both $0\nu\beta\beta$ and single-electron events were distributed uniformly throughout the detector volume.

\begin{figure}[htbp]
\centering
\includegraphics[width=0.75\textwidth]{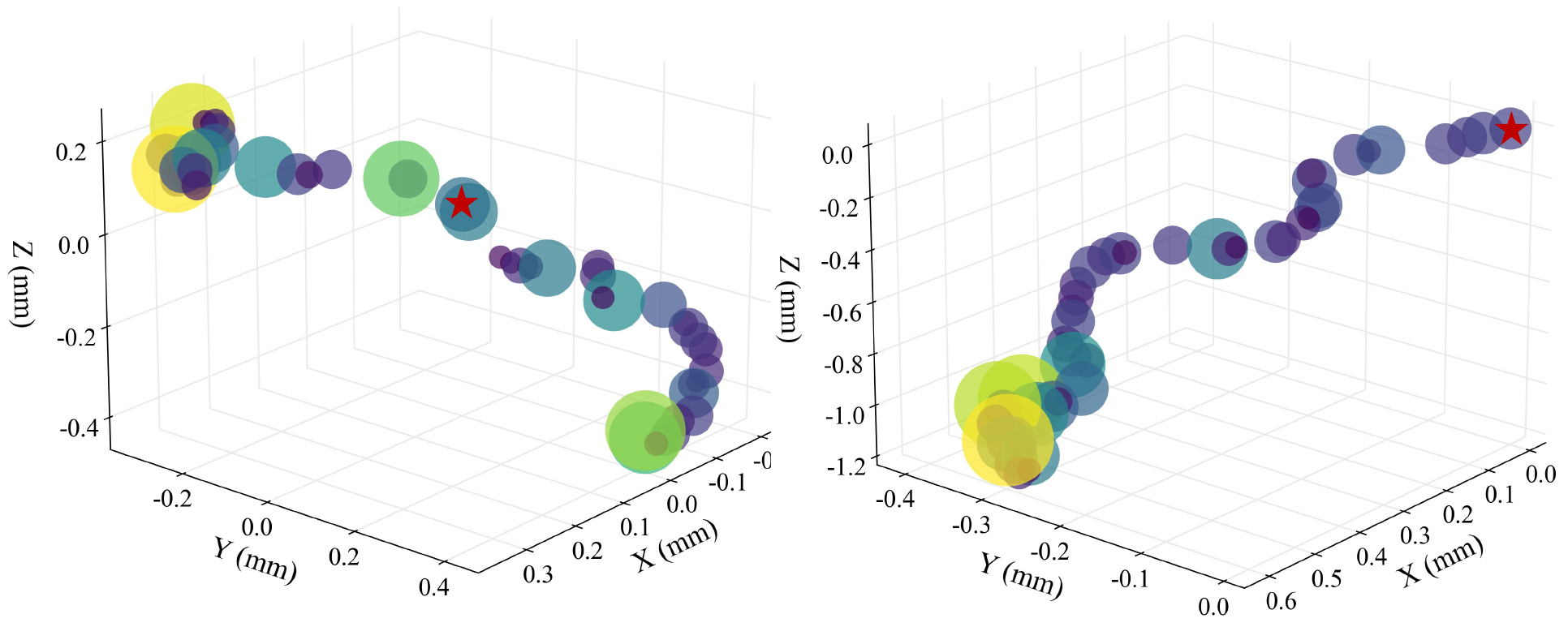}
\caption{Spatial distributions of energy deposition for a $0\nu\beta\beta$ event (left) and a single-electron event (right). The red star denotes the initial vertex at (0,0,0). Marker sizes and colors (from purple to yellow) scale with the energy deposited ($\Delta E_i$) at each interaction step.
}
\label{fig:track}
\end{figure}

Particle interactions within the germanium crystal were simulated using Geant4 (version 4.11.2)~\cite{geant4_2003}, employing the \texttt{G4EmStandardPhysics\_option4}~\cite{geant4_2016} physics list for accurate calculation of low-energy electron interactions. The particle production range cut was set to 1~$\mu$m to obtain precise tracking of electron trajectories. The output of the Geant4 simulation for each event consisted of a set of $N$ discrete interaction points, denoted as $\Omega$:
\begin{equation}
    \mathbf{\Omega} = \{ (x_i, y_i, z_i, \Delta E_i) \}_{i=1}^{N}
\end{equation}
where $(x_i, y_i, z_i)$ are the spatial coordinates and $\Delta E_i$ is the energy deposited at each interaction step. Fig.~\ref{fig:track} shows the simulated spatial distribution of these interaction points for a typical $0\nu\beta\beta$ event and a single-electron event. These data served as the input for the charge cloud modeling.

\subsection{Charge Cloud Modeling}
\label{sec2.2}
\begin{figure}[htbp]
\centering
\includegraphics[width=0.5\linewidth]{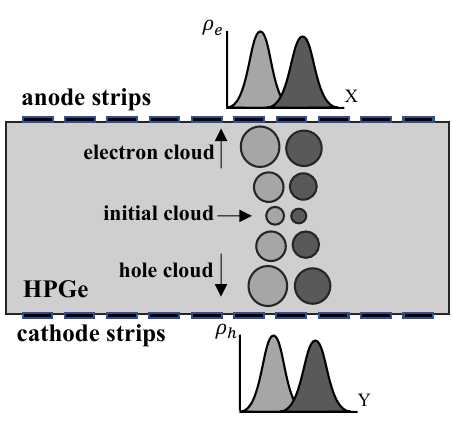}
\caption{Schematic of the charge cloud evolution and collection process in an orthogonal-strip HPGe detector. Electron and hole clouds drift toward opposite electrodes, expanding through Coulomb repulsion and thermal diffusion. When the cloud size becomes comparable to the strip pitch, charge sharing occurs across multiple electrodes. The cathode and anode strips are depicted non-orthogonally here for clarity.}
\label{fig:charge_cloud}
\end{figure}

To balance physical accuracy with computational efficiency, a hybrid numerical-analytical approach was developed to model charge cloud dynamics. For each event, the discrete interaction points obtained from Geant4 were treated as a set of macro-particles. The drift and mutual Coulomb repulsion of these macro-particles were numerically simulated, while the internal expansion of each macro-particle was modeled analytically. Figure~\ref{fig:charge_cloud} provides a schematic illustration of the charge cloud evolution, showing the combined effects of drift, Coulomb repulsion, diffusion, and the resulting charge sharing at the electrode surface. The simulation proceeded as follows:

\begin{enumerate}
    \item \textbf{Macro-particle Initialization:} For each event, the discrete interaction points $(x_i, y_i, z_i, E_{dep,i})$ obtained from Geant4 were instantiated N macro-particles (electron and hole clouds), where N was typically several tens to nearly one hundred. Each macro-particle represented a localized initial charge cloud.
    \item \textbf{Drift and Repulsion Simulation:} The trajectories of these macro-particles were simulated using SolidStateDetectors.jl. The simulation accounted for drift under the external electric field and the mutual Coulomb repulsion between different macro-particles, which governed the global expansion of the charge cloud ensemble. The final positions and drift times $(s_{f,i}, t_{drift,i})$ were recorded as macro-particles reached the electrodes (where $s=x$ for electrons collected by the top electrode and $s=y$ for holes collected by the bottom electrode).
   \item \textbf{Analytical Expansion and Charge Collection:} The internal expansion of each macro-particle was modeled analytically. Upon arrival at the electrode, the projected energy distribution of each macro-particle was approximated by a Gaussian profile. For an event with $N$ macro-particles, the total energy density $\rho(s)$ was calculated as the superposition of these distributions:
    \begin{equation}
        \rho(s) = \sum_{i=1}^{N} \frac{E_{dep, i}}{\sqrt{2\pi}\sigma_i} \exp\left( -\frac{(s - s_{f,i})^2}{2\sigma_i^2} \right)
        \label{eq:density}
    \end{equation}
   where $s_{f,i}$ is the final coordinate of the $i$-th macro-particle, and the diffusion width $\sigma_i$ is given by $\sigma_{i, e/h} = \sqrt{2 D_{e/h} t_{drift, i}}$ ($D$ is the diffusion coefficient for electrons/holes). Internal self-repulsion within each macro-particle was neglected as it is secondary to thermal diffusion~\cite{He_charge_sharing_diffonly,Du_charge_sharing_diffonly,KIM_charge_sharing_diffonly}. The energy $E_k$ collected by the $k$-th strip was obtained by integrating $\rho(s)$ over the strip boundaries $[s_k, s_{k+1}]$:
    \begin{equation}
        E_k = \int_{s_k}^{s_{k+1}} \rho(s) ds = \sum_{i=1}^{N} \frac{E_{dep,i}}{2} \left[	\text{erf}\left(\frac{s_{k+1} - s_{f,i}}{\sqrt{2}\sigma_i}\right) - \text{erf}\left(\frac{s_k - s_{f,i}}{\sqrt{2}\sigma_i}\right) \right]
        \label{eq:strip_energy}
    \end{equation}
\end{enumerate}

\begin{figure}[htbp]
\centering
\includegraphics[width=0.8\linewidth]{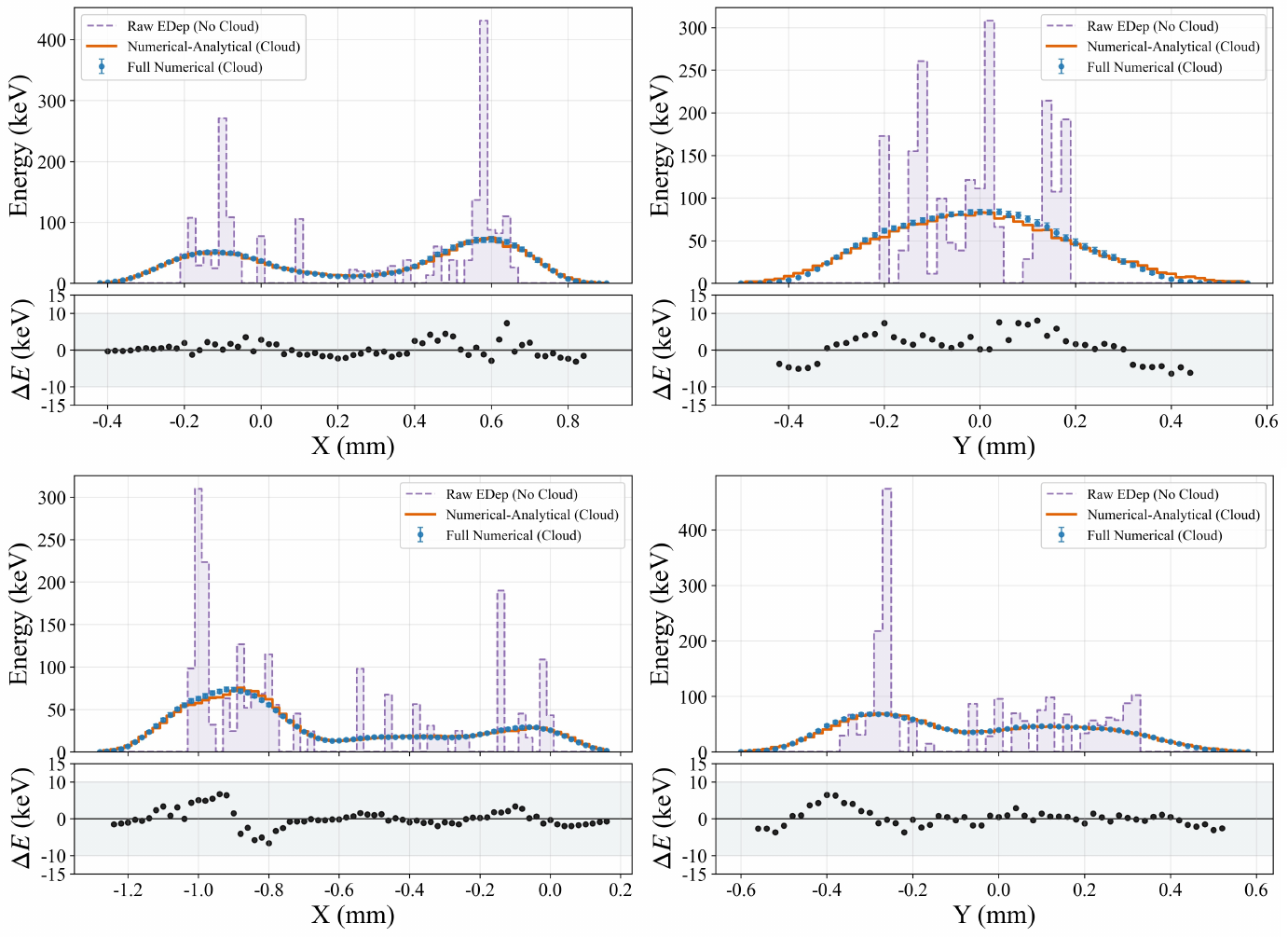}
\caption{Validation of the hybrid numerical-analytical approach against full numerical simulations. Energy profiles projected along the x-axis for electrons collected by the top electrode (left) and along the y-axis for holes collected by the bottom electrode (right). The top row shows a typical $0\nu\beta\beta$ event, while the bottom row shows a single-electron event. In each panel, the initial energy deposition from Geant4 (light purple dashed line) is compared with the hybrid approach (orange curves) and full numerical simulations (blue points with error bars representing mean and standard deviation from multiple runs). 
The relative standard deviation remains within 10\%, reflecting the statistical stability of charge collection for these MeV-scale events. Residuals (black dots) and ±10~keV bands (shaded) are shown in the lower panels.
}
\label{fig:model_validation}
\end{figure}

The hybrid numerical-analytical approach was validated against full numerical simulations performed with SolidStateDetectors.jl, which served as the benchmark in this study. As illustrated in Fig.~\ref{fig:model_validation}, the energy profiles from the hybrid approach showed good agreement with the full numerical results. The residuals between the two methods remained within 10~keV, validating the accuracy of the hybrid model in modeling the charge transport dynamics. Additionally, the computational efficiency was significantly improved, requring only a few seconds per event compared to several hours for full numerical simulations. For each simulated event, the strip readouts were recorded as two vectors $\mathbf{E}_{\text{top}}$ and $\mathbf{E}_{\text{bottom}}$, which served as input for subsequent CNN-based topological classification.

\section{Event Discrimination Method}
\label{sec3}
\subsection{Dataset Preparation and Preprocessing}
\label{sec3.1}
Datasets for both the $0\nu\beta\beta$ and single-electron events were generated using the simulation framework described in Section~\ref{sec2}. The energies collected from the orthogonal strips were extracted into two 1D vectors, $\mathbf{E}_{\text{top}}$ and $\mathbf{E}_{\text{bottom}}$. These vectors were then processed for CNN-based classification as follows:
\begin{description}[leftmargin=0pt,
  itemsep=0.em,
  before=\vspace{0.em},
  after=\vspace{0.em}]
\item [Event Selection]
To account for electronic noise, a readout energy threshold of 1~keV was applied to each strip. Only strips with deposited energy exceeding this threshold were recorded as valid hits. Events were further selected if the total reconstructed energy on both readout planes fell within the $2039\pm3$~keV window, corresponding to the region of interest (ROI) for $0\nu\beta\beta$ decay.
\item [Feature Engineering] To standardize the CNN input, the energy vectors $\mathbf{E}_{\text{top}}$ and $\mathbf{E}_{\text{bottom}}$ were mapped onto fixed-length vectors of 32 elements, which covered the maximum projected track length. The recorded energies were placed at the center of each vector, with zeros padded to the remaining positions:
\begin{equation}
\mathbf{E}_{\text{top/bottom}} = [0, \dots, 0, E_{i}, E_{i+1}, \dots, E_{j}, 0, \dots, 0]
\end{equation}
\item [Dataset Partitioning] 
A balanced subset of 35,000 events per class was randomly selected and divided into training and test sets at an 80:20 ratio.
\end{description}
The discrimination between $0\nu\beta\beta$ and single-electron events relies on their distinct topological signatures as illustrated in Fig.~\ref{fig:model_validation}. The $0\nu\beta\beta$ signal typically exhibits a characteristic two-blob topology in its energy profile, while single-electron events display a single-blob structure. These topological differences provide the physical basis for CNN-based classification.

\subsection{Classification Using Convolutional Neural Networks}
\label{sec3.2}
\begin{figure}[htbp]
    \centering
    \includegraphics[width=0.55\linewidth]{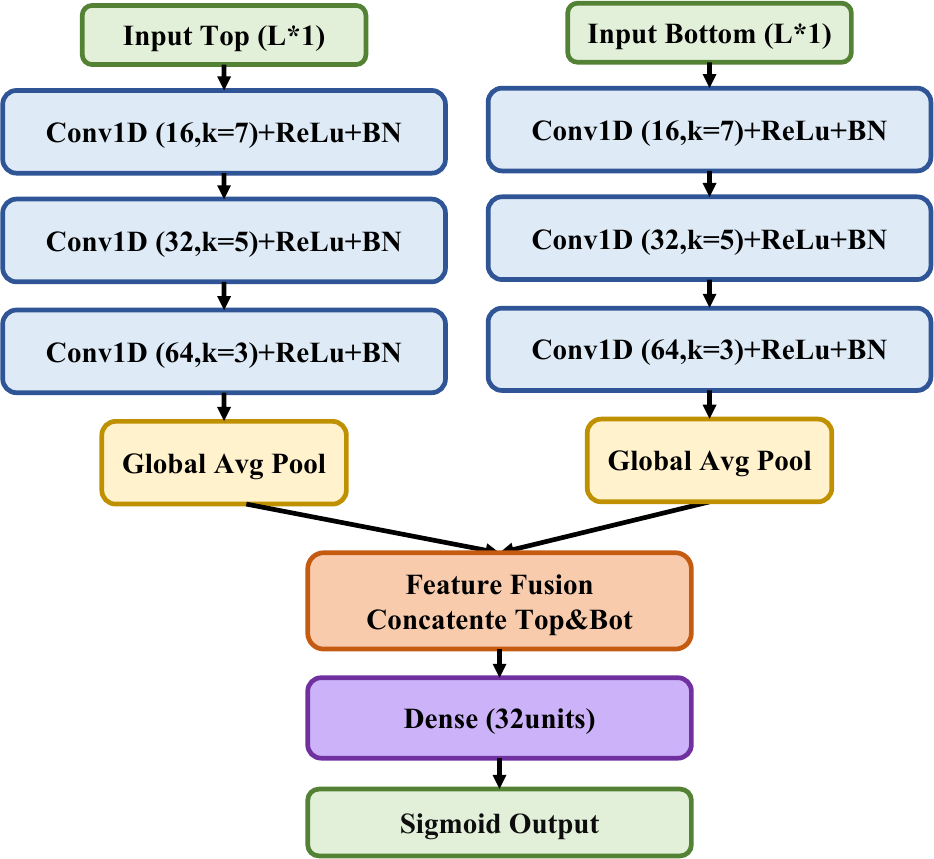} 
    \caption{The dual-branch CNN architecture for event discrimination. Independent feature extraction branches process the top and bottom strip energy vectors, followed by a fusion module that combines the extracted features for final classification.}
    \label{fig:model_workflow}
\end{figure}

A dual-branch CNN architecture was employed to classify $0\nu\beta\beta$ events from single-electron events based on the energies collected by the orthogonal strips. The network processed the independent top and bottom energy vectors through two parallel branches, as illustrated in Fig.~\ref{fig:model_workflow}. The architecture and training procedure are described as follows:
\begin{description}[leftmargin=0pt,
  itemsep=0.em,
  before=\vspace{0.em},
  after=\vspace{0.em}]
\item[Feature Extraction Branches] Each branch processed the energy vector through three convolutional layers (comprising 16, 32, and 64 filters) with kernel sizes of 7, 5, and 3, respectively. Each convolutional layer applied a ReLU activation and was followed by batch normalization. Global average pooling was then applied to aggregate the extracted topological features into compact vectors.
    
\item[Fusion Module] The resulting vectors from both branches were concatenated and fed into a fully connected layer. Dropout was applied to prevent overfitting, and a sigmoid function produced the final classification score.
    
\item[Training Procedure] The model was implemented in TensorFlow/Keras and trained with the Adam optimizer (initial learning rate of 0.0005) under a binary cross-entropy loss function. Training was monitored using an early stopping criterion with a patience of 12 epochs to mitigate overfitting. Model performance was evaluated using the area under the receiver operating characteristic curve (ROC-AUC).
\end{description}

\section{Results and Discussion}
\label{sec4}
\subsection{CNN Discrimination Performance}
\label{sec4.1}
\begin{figure}[H]
    \centering
    \includegraphics[width=0.9\linewidth]{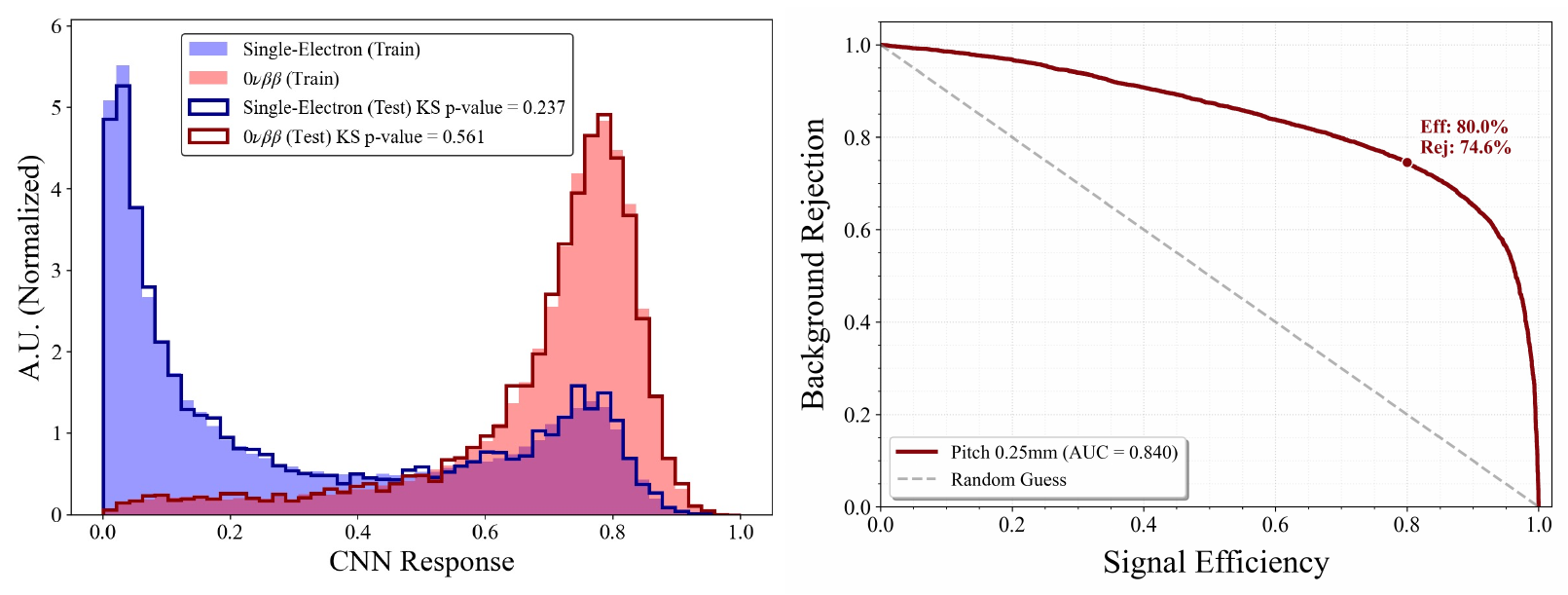}
    \caption{Discrimination performance of the dual-branch CNN for the baseline detector configuration (a 15-mm thickness and a 0.25-mm strip pitch). The left panel shows the CNN response distribution for $0\nu\beta\beta$ events (red) and single-electron events (blue) in the training and test samples, with a K-S test p-value of 0.561. The right panel shows the corresponding ROC curve with an AUC of 0.840.}
    \label{fig:cnn_base}
\end{figure}

The performance of the dual-branch CNN for the baseline detector configuration (15-mm thickness and 0.25-mm strip pitch) is presented in Fig.~\ref{fig:cnn_base}. Good separation between $0\nu\beta\beta$ events and single-electron events is observed in the network's response distributions. The Kolmogorov-Smirnov (K-S) test yields a p-value of 0.561, indicating good agreement between training and test samples and suggesting no significant overtraining. The model achieves an AUC of 0.840, corresponding to a background rejection rate of 74.6\% at a fixed signal efficiency of 80.0\%. These results confirm that the proposed CNN architecture effectively captures the topological features necessary to distinguish $0\nu\beta\beta$ events from single-electron events.

\subsection{Impact of Strip Pitch on Discrimination}
\label{sec4.2}
\begin{figure}[htbp]
    \centering  
    \includegraphics[width=0.75\linewidth]{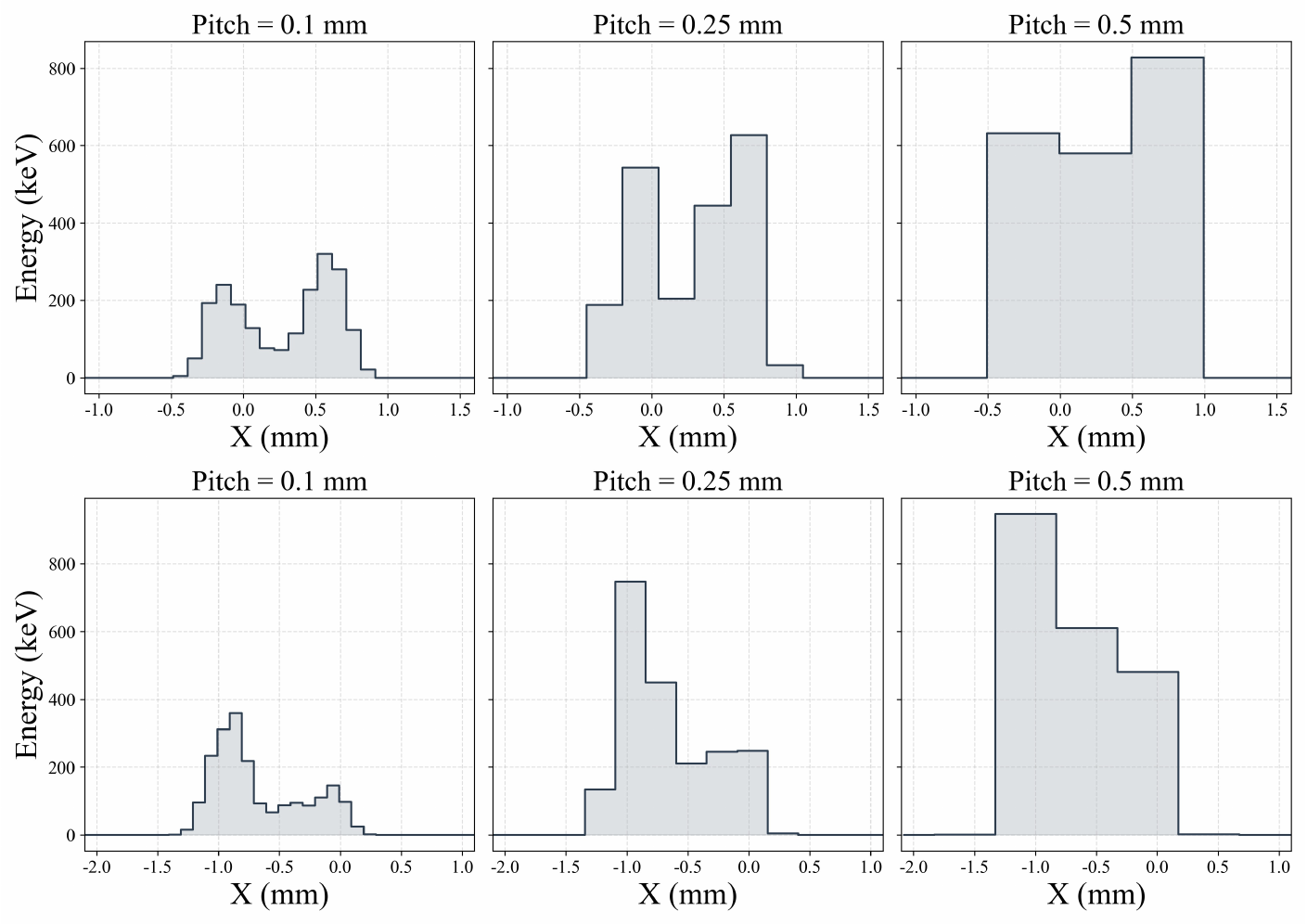} 
    \caption{Comparison of energy profiles projected along the x-axis for $0\nu\beta\beta$ events (top row) and single-electron events (bottom row) at strip pitches of 0.1, 0.25, and 0.5 mm, with a fixed crystal thickness of 15 mm.}
    \label{fig:diffpitch_energy_profile}
\end{figure}

The strip pitch governs the spatial resolution, which determines the capability to extract topological features of $0\nu\beta\beta$ events. Figure~\ref{fig:diffpitch_energy_profile} illustrates how strip pitch affects the topology of $0\nu\beta\beta$ and single-electron events. At fine pitches (0.1 and 0.25~mm), $0\nu\beta\beta$ events exhibit a characteristic two-blob structure, while single-electron backgrounds show a distinct single-blob pattern. However, at a coarser pitch of 0.5~mm, these topological features undergo significant spatial blurring, thereby degrading the discriminating power.

\begin{figure}[htbp]
    \centering
    \includegraphics[width=0.54\linewidth]{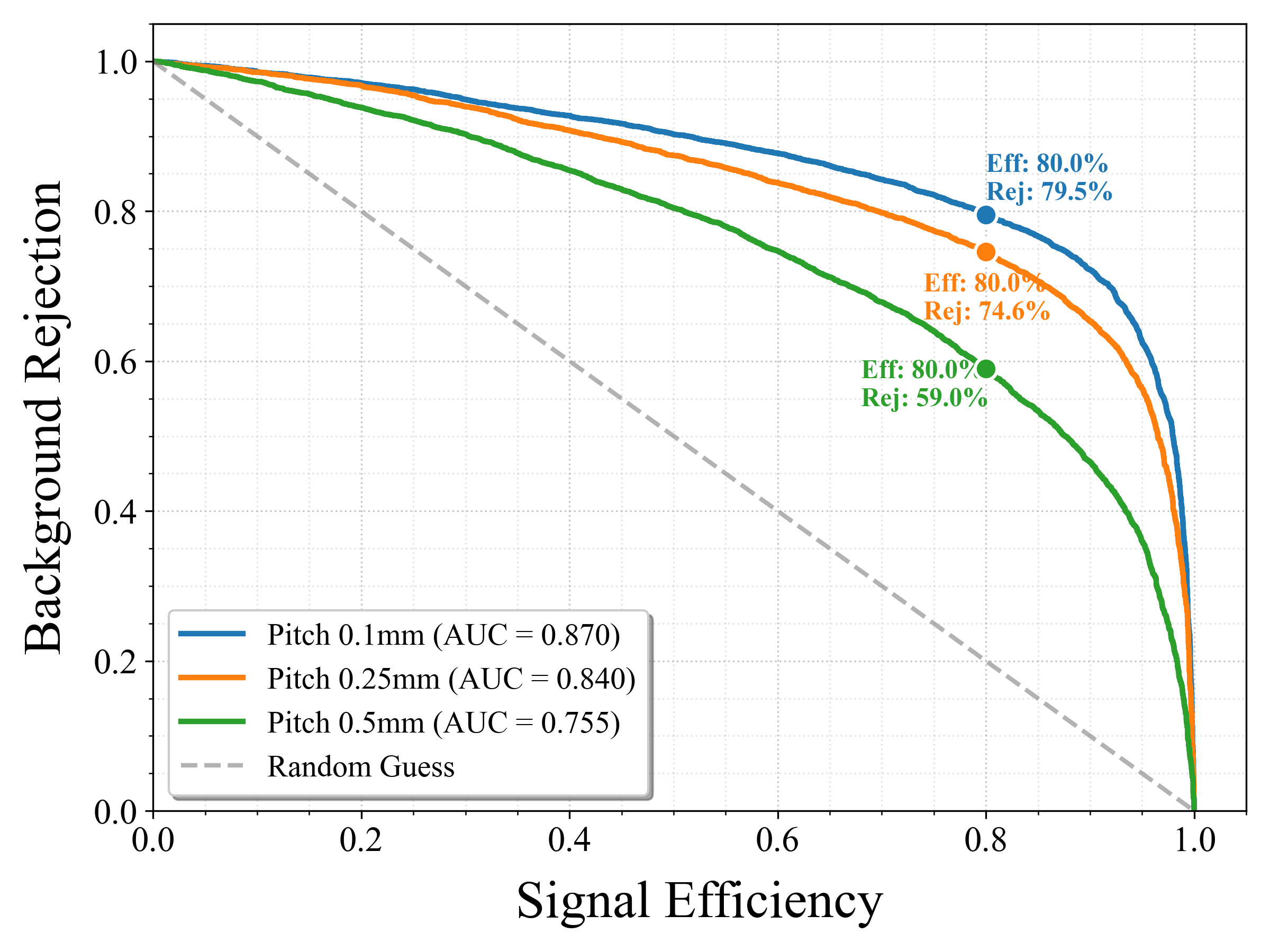} 
    \caption{Impact of strip pitch on background discrimination performance. ROC curves are compared for strip pitches ranging from 0.1~mm to 0.5~mm, with a fixed crystal thickness of 15~mm.}
    \label{fig:roc_pitch}
\end{figure}

The performance degradation is quantified by the ROC curves shown in Figure~\ref{fig:roc_pitch} for a fixed crystal thickness of 15~mm. As the strip pitch increases from 0.1~mm to 0.5~mm, the AUC decreases from 0.870 to 0.755. Specifically, at a fixed signal efficiency of 80.0\%, the background rejection rate drops from 79.5\% (at 0.1~mm) to 59.0\% (at 0.5~mm). These results demonstrate that high-granularity electrode segmentation is essential for preserving the topological features required for effective event identification.

\subsection{Impact of Crystal Thickness on Discrimination}
\label{sec4.3}
\begin{figure}[H]
    \centering
    \includegraphics[width=0.7\linewidth]{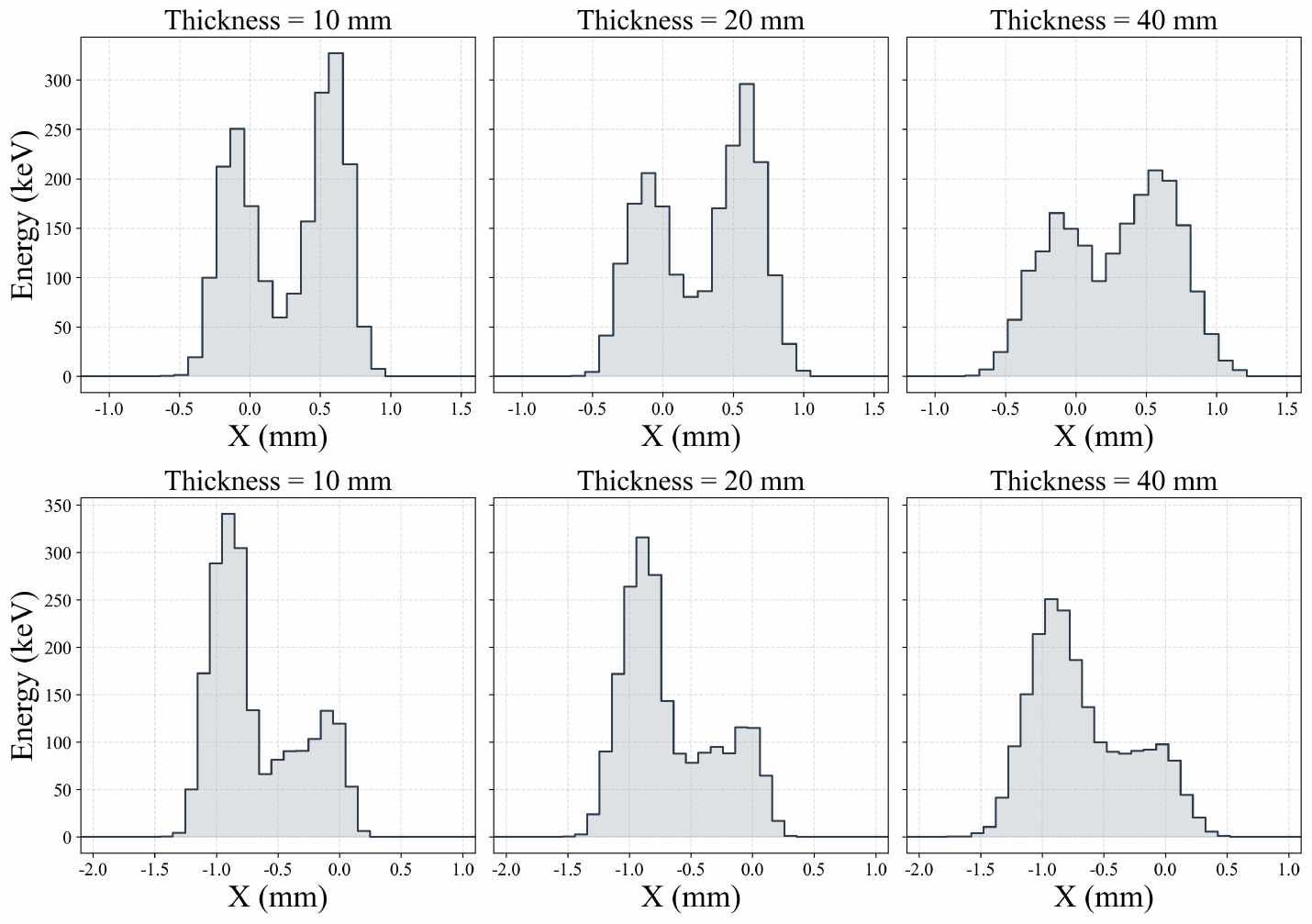}
    \caption{Comparison of energy profiles projected along the 
x-axis for a $0\nu\beta\beta$ event (top row) and a single-electron event (bottom row) for crystal thicknesses of 10, 20, and 40 mm, with a fixed strip pitch of 0.1 mm.}
\label{fig:cloud_diffheight}
\end{figure}

The crystal thickness affects both the FEP efficiency and the charge cloud broadening. Thicker crystals enhance FEP efficiency but at the cost of longer carrier drift times. The resulting charge cloud broadening obscures the topological features essential for signal discrimination. Figure~\ref{fig:cloud_diffheight} illustrates the evolution of energy profiles as a function of crystal thickness. For a thin 10-mm crystal, both $0\nu\beta\beta$ and single-electron events maintain their distinctive topologies. However, as the thickness increases to 40~mm, the characteristic two-blob structure of $0\nu\beta\beta$ events becomes obscured, making discrimination from single-electron events increasingly difficult.

\begin{figure}[H]
    \centering
    \includegraphics[width=0.95\linewidth]{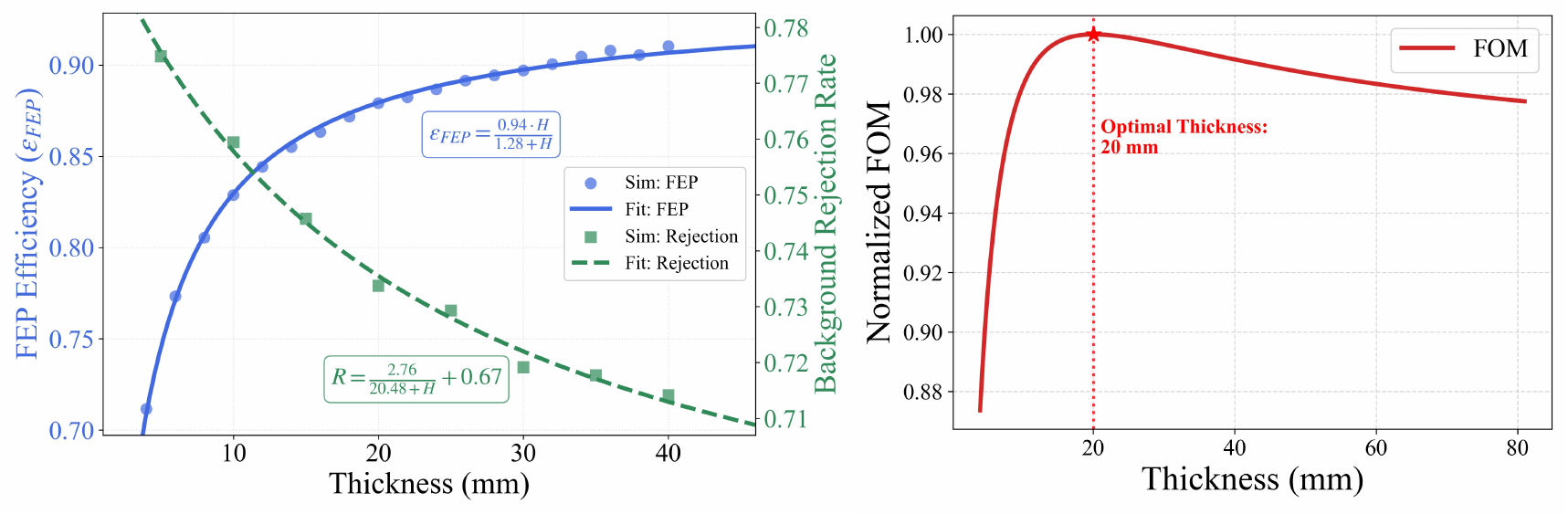}
    \caption{Optimization of crystal thickness based on experimental sensitivity. Left: FEP efficiency($\epsilon_{\text{fep}}$) and background rejection rate (at 80.0\% signal efficiency) as a function of crystal thickness. Right: The FOM calculated using Eq.~\ref{eq:fom}, identifying 20~mm as the optimal thickness.}
    \label{fig:fom_thickness}
\end{figure}

To determine the optimal thickness that balances the FEP efficiency against background suppression capability, a Figure of Merit (FOM) is defined based on the experimental half-life sensitivity:
\begin{equation}
T_{1/2} \propto \frac{\epsilon_{\text{fep}}}{\sqrt{1 - \eta_{\text{rej}}}}
\label{eq:fom}
\end{equation}
where $\epsilon_{\text{fep}}$ is the FEP efficiency and $\eta_{\text{rej}}$ is the background rejection rate. As shown in Figure~\ref{fig:fom_thickness}, increasing crystal thickness improves FEP efficiency while reducing background rejection. The FOM reaches its maximum at 20~mm, indicating the optimal thickness for maximizing experimental sensitivity.

\section{Conclusions}
\label{sec5}
In this work, a simulation framework was developed to evaluate the capability of orthogonal-strip HPGe detectors to distinguish $0\nu\beta\beta$ events from single-electron events. Particle interactions were simulated using Geant4, followed by a hybrid numerical-analytical approach that modeled charge cloud dynamics. A dual-branch CNN was employed to extract topological features from the orthogonal strip energy profiles, achieving effective event classification on the simulated datasets.

The influence of detector geometry on discrimination performance was quantitatively investigated. Strip pitch was identified as the dominant factor governing track topology resolution, with background rejection degrading significantly as pitch increased from 0.1 mm to 0.5 mm. Crystal thickness was found to exhibit a trade-off between FEP efficiency and background suppression. A crystal thickness of 20 mm is recommended for detectors with 0.25-mm strip pitch. These findings provide essential quantitative guidance for the design of advanced HPGe detectors in the field of rare-event physics experiments.

\acknowledgments
This work was supported by the National Natural Science Foundation of China (Grants No. U1865205).

\bibliographystyle{JHEP}
\bibliography{./reference}
\end{document}